# A New General Theory of Superconductivity

R. Riera, J.L. Marín and R. Betancourt-Riera
Departamento de Investigación en Física
Universidad de Sonora, Apartado Postal 5-088, 83190
Hermosillo, Sonora, MEXICO

A general theory of superconductivity based on the pairing of electrons that belong to different electronic bands is presented. These electronic bands arise because of a symmetry breaking at the critical temperature in such a way that one electron sees the other as if its mass would be negative. This symmetry breaking leads also to an energy gap (centered round of the Fermi level) between the coupled electrons. Under the assumption that electric resistance is due to electron-phonon interaction, a general equation from which the fundamental properties of any metallic superconductor can be derived is also presented. In the frame of this theory, the electron-phonon interaction (or electric resistance) is cancelled out by the electron-electron interaction in the super-conducting state. Moreover, the Bardeen-Cooper-Schrieffer (BCS) theory is a particular case of our theory and equations, a matter, which leads to a better overall agreement with experimental data.

## I. INTRODUCTION

A well known distinctive property of the superconductors is the jump of their electric resistance to zero as they arrive its' critical temperature. It is also well known that the electric resistance in metallic conductors is determined by the electron-phonon interaction (EPI). This means that when a metallic conductor becomes a superconductor, the electron-phonon interaction should be eliminated or compensated by another interaction. The only reasonable way to eliminate the EPI is to form a pair of electrons, in which the Coulomb repulsion allows that the electron-phonon interaction be compensated for certain temperature and thus the electric resistance of the material can be converted to zero. Taking into account this argument, the superconductivity in metallic conductors could not exist without the formation of such electron pairs as it takes place in the case of the BCS theory[1].
The BCS theory[1] assumes that in the superconducting state the Coulomb repulsion plus electron-phonon interaction is a negative and constant number. This means that the electron-phonon interaction is predominant with respect to the Coulomb interaction and therefore, the electric resistance would remain and the superconductor transition cannot occur. The instability of the Cooper pairs persists even if the negative potential is very weak. This is a very remarkable difference from the usual two-body problem. If we have only two particles coupled by an attractive interaction of finite range, they would not form bound states unless the attractive interaction exceeds a certain threshold.
It is then clear that in metallic conductors, the superconductivity should occur when in the coupled electron pairs the electron-phonon interaction is eliminated by the Coulomb repulsion. Within the BCS theory it means that the attractive potential is zero and that it is unnecessary that the electron-phonon interaction be greater than the Coulomb repulsion.
The phenomenon of superconductivity was discovered at the beginning of the past century (1911). Although the development of the microscopic theory of superconductivity involved a great deal of efforts, it was nevertheless finalized about forty years later (1957). In spite of this, the extensive investigations are still being carried out in the field of superconductivity, let alone its applications in engineering, and the range of topics brought under the purview of research in superconductivity is being continuously extended. Also new fields like the creation and study of organic and heavy-fermion superconductors emerged around the decade of 1970. The studies of superconductors in non-equilibrium states, magnetic superconductors, organic superconductors and ceramic superconductors were initiated somewhat earlier and become of extremely importance at the moment. However, the problem of high-temperature superconductivity in ceramic superconductors is of utmost importance and we would even call it the number one question in superconductivity.
The BCS theory in its original and simplest form enables us, even in a first approximation, to see why the critical temperature of a common metal is not high. Then one can outline possible ways in which critical temperature can, in principle, be substantially increased. The fundamental equation that constitutes the dispersion law of the pairs in the BCS theory[1] was not correctly interpreted and used to obtain the physical properties that characterize a superconductor. The limit in the critical temperature is due fundamentally to the relationship between the energy gap of the pairs (the binding energy) and the critical temperature is fixed in 3.50 without a physical reason.

The BCS theory[1] of the superconductivity can be very well called a theory of the critical values. It is useful only for superconductors characterized by a critical temperature of less than few Kelvin. From this theory one can obtain the fundamental properties of the superconductors of very low critical temperature, like the gap energy, the critical magnetic field, the jump in the specific heat, etc. This last equation is based on the derivative of the gap energy square with respect to the temperature evaluated in the critical temperature, which is incorrect, as we will see later). However, the variations of these magnitudes with the temperature as are: the gap energy with the temperature, the critical magnetic field, the energy, the entropy and the variation of the heating capacity with the temperature are not obtained correctly with the BCS theory[1].

On the other hand, the BCS theory[1], in their fundamental equation in order to obtain the wave function and the energy of the pairs, doesn't take into account the temperature in the Hamilton operator. From their equation it is possible to obtain only one part of the solution, the negative half of the gap, while the positive half is extrapolated. To consider the temperature in the initial Hamilton operator is a very difficult task. We should keep in mind the breakup of the symmetry in the transition temperature for arrive the correct solution.

We propose in this paper to make a general theory of the superconductivity worth for metallic superconductors of high or low temperature either. The equations are generally obtained by starting from properties of symmetry and then considering that in the temperature of transition the electron-phonon interaction is annulled by the dispersion of Coulomb like the only form of getting zero electric resistance. The results of the BCS theory[1] at the critical temperature and also at zero Kelvin are obtained as particular cases of our equations and contributed the equations for these properties in the interval of temperature between zero and the critical temperature. We also contributed with physical interpretation that makes it logical and flow from the foundations of our theory.

## II. PHYSICAL MODEL

Let assume that the superconductivity is due to pairs of electrons that don't feel the electric resistance of the metal upon arriving to the transition temperature, $T_C$. These pairs are formed with electrons that are slightly above and below the Fermi level, with opposed momenta $(\vec{k},-\vec{k})$ and are changing permanently their state from the $(\vec{k},-\vec{k})$ to the $(\vec{k}+\vec{q},-\vec{k}-\vec{q})$ one with the emission and absorption of a virtual phonon of momentum $\vec{q}$. That is to say, they pass from the $(\vec{k},-\vec{k})$ state to the $(\vec{k}+\vec{q},-\vec{k}-\vec{q})$ one, conserving the energy and the momentum and vice versa. We also suppose that there exists a maximum of the binding energy of the pair $2\varepsilon_0$ and that any overflow of this limit energy leads to the pair breaking and the superconductivity will be lost. This maximum energy depends only on the critical temperature, which is a characteristic of any superconductor material.

Fig. 1 shows a graph of the energy of electrons $\varepsilon(\vec{k})$ forming a pair as a function of the momentum $\vec{k}$. Notice that there no exists any band of the parabolic type that passes simultaneously by the points $1(-\vec{k}-\vec{q},\varepsilon), 1'(-\vec{k},\varepsilon+2\varepsilon_0), 2(\vec{k}+\vec{q},\varepsilon+2\varepsilon_0)$ and $2'(\vec{k},\varepsilon)$. The dashed lines correspond to the phonons and the solid lines to the electron transitions due to the emission and absorption of phonons. The electron in the positions 1' and 2 always emits phonons, while in the positions 1 and 2' absorbs phonons.

In the Fig. 2 we show that for the electrons to form a pair it is necessary they belong to different bands. Therefore, one of the bands is inverted with respect to the Fermi level, behaving like a band of electrons of negative mass. In this situation, the electron 2 emits a phonon that is absorbed by the electron 1, therefore an electron changes from state 2 to the 2' one, and the other electron changes from 1 to the 1' state. Then the electron 1' emits a phonon that is absorbed by the 2', and in this case an electron changes from state 1' to the 1, and the other electron changes from 2' to the 2, and this chain of transitions will be repeated indefinitely. The exchange of the electrons by the phonons is a virtual process. In fact the lattice participates in the process as is shown at the Fig. 3.

Notice that the states above $\varepsilon_2$ and below $\varepsilon_1$ are occupied, thus the simultaneous existence of conductivity and superconductivity is forbidden. We should take into account that only the conductivity by simple electrons can exist simultaneously with superconductor pairs, when there is a magnetic field and the electronic state are inside of the zone determined by $2\varepsilon_0$.

From Fig. 2 we can obtain the characteristic equations of the electronic bands:

$$\varepsilon_1 = -\frac{\hbar^2 k^2}{2m^*} + \varepsilon_F \text{ and } \varepsilon_2 = \frac{\hbar^2 k^2}{2m^*} - \varepsilon_F \tag{1}$$

These equations only have in common the energy of Fermi $\varepsilon_F$ and the momentum at the Fermi surface $\vec{k}_F$. Combining them we obtain

$$\varepsilon_F = \frac{\hbar^2 k_F^2}{2m^*} \text{ where } k_F = \pm\sqrt{\frac{m^*}{\hbar^2}\varepsilon_F} \tag{2}$$

From Fig. 2 we can also find that

$$-(k+q) < -k_F < -k \text{ and } k < k_F < k+q \tag{3}$$

and then

$$k_F = k + \frac{q}{2} \tag{4}$$

### III. CONSIDERATIONS FROM THE TWO BODY PROBLEM

Let us introduce the transformations

$$\vec{r} = \vec{r}_2 - \vec{r}_1, \quad \vec{R} = \frac{m_1 \vec{r}_1 + m_2 \vec{r}_2}{M}, \quad M = m_1 + m_2 \tag{5}$$

$$\vec{r}_{12} = \vec{R} \pm \frac{m_{21}}{M}\vec{r}, \quad \mu = \frac{m_1 m_2}{M} \tag{6}$$

which allow us to separate the problem in

Particle 1 center of mass $\vec{R} = \frac{m_1 \vec{r}_1 + m_2 \vec{r}_2}{M}, \quad \vec{P} = \vec{p}_1 + \vec{p}_2, \quad \vec{L} = \vec{R} \times \vec{P}$ (7)

Particle 2 reduced mass $\vec{r} = \vec{r}_2 - \vec{r}_1, \quad \vec{p} = \frac{m_2 \vec{p}_1 - m_1 \vec{p}_2}{M}, \quad \vec{l} = \vec{r} \times \vec{p}$ (8)

Being

$$\frac{p_1^2}{2m_1}+\frac{p_2^2}{2m_2}=\frac{p^2}{2\mu}+\frac{P^2}{2M}, \quad \vec{l}_1+\vec{l}_2=\vec{l}+\vec{L}, \quad \vec{l}_1=\vec{r}_1\times\vec{p}_1, \quad \vec{l}_2=\vec{r}_2\times\vec{p}_2 \qquad (9)$$

Observations:

The electrons forming a pair have $\vec{k}_1=-\vec{k}_2$, which implies that $\vec{p}_1=-\vec{p}_2$

1. The center of mass of the pair doesn't move, that is in a good coincidence with the idea that the pair is located in $\vec{P}=0$, and $\vec{L}=0$.

2. The particle of reduced mass has momentum $\pm\vec{k}_1=\pm\vec{k}_2$ and angular momentum $\vec{l}=\pm 2\vec{l}_1=\pm 2\vec{l}_2$, besides their kinetic energy oscillates between the energies $\hbar^2 k_1^2/2\mu$ and $-\hbar^2 k_2^2/2\mu$, thus keeping constant the total energy.

## IV. GENERAL EQUATION OF THE SUPERCONDUCTOR PAIR

From Fig. 2 or 4 we can see that for paired electrons there is a constant energy $E$, which is related to energies $\varepsilon$ and $\varepsilon_0$ through a relationship similar to that of the sides in a rectangular triangle.

We will keep in mind from the same graph that

$$\varepsilon(k+q)-\varepsilon(-k-q)=\varepsilon(k)-\varepsilon(-k) \qquad (10)$$

and all the energies are taken relative to the level of Fermi, then

$$\varepsilon(k)=\frac{\hbar^2 k^2}{2m^*}-\varepsilon_F \text{ and } \varepsilon(-k)=-\frac{\hbar^2 k^2}{2m^*}+\varepsilon_F \qquad (11)$$

Therefore

$$\varepsilon(k)-\varepsilon(-k)=\frac{\hbar^2 k^2}{m^*}-2\varepsilon_F=2\varepsilon(k) \qquad (12)$$

Keeping in mind that the relationship between any of the triangles 1'1"2' and 12"2 is equal, we arrived to the following equation

$$E'(k)=\sqrt{(\varepsilon(k)-\varepsilon(-k))^2+(2\varepsilon_0)^2} \qquad (13)$$

then

$$E'(k)=2\sqrt{\varepsilon^2(k)+\varepsilon_0^2} \qquad (14)$$

This energy represents the maximum energy that it is necessary to give (Kelvin degree starting from zero) in order to break the pair and it is a constant. This equation is the fundamental one of the superconductor pairs.

## V. THE HAMILTONIAN OPERATOR OF THE ELECTRON SYSTEM

The Hamilton operator of electron system above the critical temperature can be written in the following form

$$H_e=H_T+H_{ee}+H_{eP} \qquad (15)$$

where $H_e$ is the total electron energy, $H_T$ is the kinetic energy, $H_{ee}$ is the Coulomb repulsion, and $H_{eP}$ is the electron-phonon interaction.

Upon arriving to the critical temperature $H_{ee}+H_{eP}=0$ since electric resistance in the superconducting state doesn't exist. From this condition one can conclude that the electron-phonon interaction and the Coulomb electron-electron interaction depend on the temperature. It is possible to obtain from this relation the critical temperature as a function of the constants that characterize both interactions, and the deformation potential, the static dielectric constant, the effective masses of the ions and electrons and maybe also of the numbers atomic. This way, one can obtain the important condition in order to achieve materials with critical temperatures majors.

## VI. SUPERCONDUCTIVITY THEORY

The state of electrons in the pair can be described by $|\vec{k},\sigma\rangle$ and $|\vec{k}',\sigma'\rangle$, where $\vec{k}' = -\vec{k}$ is the wavevector and $\sigma = -\sigma'$ is the spin. The reduced wavevector is not a good quantum number for pairs, therefore their state should be specified by the temperature $|n, T_n\rangle$, where $n$ is an integer number, which exists if the pair takes the thermal energy in discrete form and doesn't exist if the energy is take in continuous form.

The maximum wavevector of the individual electrons corresponds to the critical temperature ($\vec{k}_{max} \rightarrow T_C$) of the pairs and the wavevector zero corresponds to zero degrees of absolute temperature ($\vec{k} = 0 \rightarrow T = 0^0 K$). The pair oscillates between the electron states $|\vec{k}\uparrow\rangle$, $|-\vec{k}\downarrow\rangle$ and $|\vec{k}+\vec{q}\uparrow\rangle$, $|-\vec{k}-\vec{q}\downarrow\rangle$, where $\vec{q}$ is the wavevector corresponding to a virtual phonon.

Let denote by $\hat{a}^+_{\vec{k}\uparrow}$ and $\hat{a}_{\vec{k}\uparrow}$ the operators of creation and annihilation of the electrons that form a pair in the state $\vec{k}$ with spin up. These operators satisfy the usual commutation relations:

$$[\hat{a}_{\vec{k}\sigma}, \hat{a}^+_{\vec{k}'\sigma'}]_+ = \delta_{kk'}\delta_{\sigma\sigma'} \tag{16}$$

$$[\hat{a}_{\vec{k}\sigma}, \hat{a}_{\vec{k}'\sigma'}]_+ = [\hat{a}^+_{\vec{k}\sigma}, \hat{a}^+_{\vec{k}'\sigma'}]_+ = 0 \tag{17}$$

The number operator of single particle $\hat{n}_{\vec{k}\sigma}$ is defined as

$$\hat{n}_{\vec{k}\sigma} = \hat{a}^+_{\vec{k}\sigma}\hat{a}_{\vec{k}\sigma} \tag{18}$$

The electrons in the pairs should have the same value of the wavevector but with opposed sign because they belong to two different bands (see Fig. 1) and they are located one above and the other below the Fermi energy, which is produced by the symmetry breakdown. Thus we can see that the pair is made up of a dynamic state that oscillates between the states

$(\vec{k}\uparrow, -\vec{k}\downarrow)$ and $(\vec{k}+\vec{q}\uparrow, -\vec{k}-\vec{q}\downarrow)$

Then

$$\hat{a}^+(-\vec{k}\downarrow)\hat{a}(-\vec{k}-\vec{q}\downarrow)\hat{a}^+(\vec{k}\uparrow)\hat{a}(\vec{k}+\vec{q}\downarrow) \tag{19}$$

Conservation of momentum requires that

$$\vec{k}+\vec{q}+(-\vec{k}-\vec{q}) = \vec{k}+(-\vec{k}) = 0 \tag{20}$$

Then we start by considering a reduced problem in which the configurations are included with the states occupied by the pairs of the two electrons with $\vec{k}\uparrow$ and $-\vec{k}\downarrow$. A pair is designated by the reduced wavevector $\vec{k}$, independent of spin. Creation and annihilation operators for pairs may be defined in terms of single electron operators as follows (see Fig 1):

$$\hat{b}_{\vec{k}}(2\varepsilon_0) = \hat{a}_{-\vec{k}\downarrow}(\varepsilon_0)\hat{a}_{\vec{k}\uparrow}(-\varepsilon_0) \tag{21}$$

$$\hat{b}_{\vec{k}}^+(2\varepsilon_0) = \hat{a}_{\vec{k}\uparrow}^+(-\varepsilon_0)\hat{a}_{-\vec{k}\downarrow}^+(\varepsilon_0) \tag{22}$$

These operators satisfy the commutation relations

$$\left[\hat{b}_{\vec{k}},\hat{b}_{\vec{k}}^+\right]_\pm = \hat{a}_{-\vec{k}\downarrow}(\varepsilon_0)\hat{a}_{\vec{k}\uparrow}(-\varepsilon_0)\hat{a}_{\vec{k}\uparrow}^+(-\varepsilon_0)\hat{a}_{-\vec{k}\downarrow}^+(\varepsilon_0) \pm \hat{a}_{\vec{k}\uparrow}^+(-\varepsilon_0)\hat{a}_{-\vec{k}\downarrow}^+(\varepsilon_0)\hat{a}_{-\vec{k}\downarrow}(\varepsilon_0)\hat{a}_{\vec{k}\uparrow}(-\varepsilon_0) \tag{23}$$

$$\left[\hat{b}_{\vec{k}},\hat{b}_{\vec{k}}^+\right]_\pm = \left[(1-n_{\vec{k}\uparrow})(1-n_{-\vec{k}\downarrow}) \pm n_{\vec{k}\uparrow}n_{-\vec{k}\downarrow}\right]\delta_{\vec{k}\uparrow,-\vec{k}\downarrow} \tag{24}$$

Considering that the single electrons $n_{-\vec{k}\downarrow}$ and $n_{\vec{k}\uparrow}$ can only take the values 0 and 1, the commutation relations are given by

$$\left[\hat{b}_{\vec{k}},\hat{b}_{\vec{k}}^+\right]_+ = \delta_{\vec{k}\uparrow,-\vec{k}\downarrow} \tag{25}$$

$$\left[\hat{b}_{\vec{k}},\hat{b}_{\vec{k}}^+\right]_- = 1 - n_{\vec{k}\uparrow} - n_{-\vec{k}\downarrow} \tag{26}$$

$$\left[\hat{b}_{\vec{k}},\hat{b}_{\vec{k}}\right]_\pm = \left[\hat{b}_{\vec{k}}^+,\hat{b}_{\vec{k}}^+\right]_\pm = 0 \tag{27}$$

The reduced wavevector $\vec{k}$ is not a good quantum number for the pairs, however the temperature is the suitable physical magnitude for the characterization of the state of the pairs for what we will consider $\hat{b}_{\vec{k}}(2\varepsilon_0) = \hat{b}_{n,T_n}$ and $\hat{b}_{\vec{k}}^+(2\varepsilon_0) = \hat{b}_{n,T_n}^+$. Then these commutation relations are transformed as

$$\left[\hat{b}_{n,T_n},\hat{b}_{n',T_n'}^+\right]_+ = \delta_{nn'}\delta_{T_nT_n'} \tag{28}$$

$$\left[\hat{b}_{n,T_n},\hat{b}_{n',T_n'}\right]_- = \left[\hat{b}_{n,T_n}^+,\hat{b}_{n',T_n'}^+\right]_- = 0 \tag{29}$$

$$\left[\hat{b}_{n,T_n},\hat{b}_{n',T_n'}\right]_+ = 2\hat{b}_{n,T_n}\hat{b}_{n',T_n'}(1-\delta_{T_nT_n'})(1-\delta_{nn'}) \tag{30}$$

$$\left[\hat{b}_{n,T_n}^+,\hat{b}_{n',T_n'}^+\right]_+ = 2\hat{b}_{n,T_n}^+\hat{b}_{n',T_n'}^+(1-\delta_{T_nT_n'})(1-\delta_{nn'}) \tag{31}$$

The last two commutation relations are always zero because one cannot simultaneously to create pairs in the same state at the same temperature or different states with different temperatures. On the one hand a distribution of pairs for states of the fixed temperature doesn't exist. On the other hand, for any temperature between $0 \leq T \leq T_C$, there only exists one pair for each unit cell, i.e. each cell of crystal contributes with one pair that has the same state determined by the temperature.

The commutation rules obtained above tell us that the pairs, in spite of the fact that they have spin zero, do not obey the bosons statistics and their behavior is more typical for the fermions. However, the distribution of them in all the allowed states between 0 and $T_C$ doesn't exist, as it is typical for fermions. The reason is that the pairs are quasi-particles that possess statistical properties, which are mixed between bosons and fermions, i.e., it is necessary to define a new statistics for them.

From commutation relationships we can arrive to

$$\hat{b}_{nT}^2 = \hat{b}_{nT}^{+2} = 0 \tag{32}$$

If $|n=0, T=0K\rangle$ is the wavefunction corresponding to the ground state then

$$\hat{b}_{nT}|n=0, T=0K\rangle = 0 \tag{33}$$

If $|n=n_{max}, T_C\rangle$ is the wavefunction corresponding to the state in $T_C$ then

$$\hat{b}_{nT}^+|n=n_{max}, T_C\rangle = 0 \tag{34}$$

The 1 and 2 relations mean that when $T_C \to 0K$ the binding energy of the pair is created and in case of $0K \to T_C$ it is annihilated (see Fig. 4). Also $\hat{b}^+_{nT}\hat{b}_{nT} = N_{nT}$ and $N^2_{nT} = N_{nT}$, which determines that the numbers of occupation of the pairs are equal to 0 or 1.

### A. Hamiltonian of the pairs

It is clear that in the case of the superconductivity the formation of the pairs occurs at the critical temperature. At this temperature the electron phonon interaction and the Coulomb repulsion are compensated each other, and the electron-phonon interaction and the Coulomb repulsion depend, both or one of them, on the temperature. With $T = T_C$ the binding energy is almost zero and alll the kinetic energy is due to the reduced wavevector of the pair or of the individual electrons. In this case this energy and the reduced wavevector are maxima.
Then

$$\hat{H} = 2\sum_{\vec{k}} \sqrt{\varepsilon^2_{\vec{k}}} \hat{b}^+_{\vec{k}} \hat{b}_{\vec{k}} = 2\sum_{\vec{k}} \sqrt{\varepsilon^2_{\vec{k}}} \hat{a}^+_{\vec{k}\uparrow}\hat{a}^+_{-\vec{k}\downarrow}\hat{a}_{-\vec{k}\downarrow}\hat{a}_{\vec{k}\uparrow} \tag{35}$$

what it is the same

$$\hat{H} = 2\sum_{\vec{k}} \sqrt{\varepsilon'_{\vec{k}} \varepsilon'_{-\vec{k}}} \hat{n}_{\vec{k}} \hat{n}_{-\vec{k}} \tag{36}$$

where

$$\hat{N}_{\vec{k}T_C} = \hat{n}_{\vec{k}} \hat{n}_{-\vec{k}} \tag{37}$$

and

$$\varepsilon^2_{\vec{k}} = \varepsilon'_{\vec{k}} \varepsilon'_{-\vec{k}} \tag{38}$$

In the case when $T < T_C$, part of the energy converts to internal energy state of the pair, then the binding energy increases and the sum of this energy with the reduced kinetic energy is equal to the maximum initial energy in $T = T_C$.

Considering that there is no interaction between the pairs and that $\varepsilon_0(T)$ is the bounding energy, the Hamiltonian can be written as

$$\hat{H} = 2\sum_{\vec{k}} \sqrt{\varepsilon^2_{\vec{k}} + \varepsilon^2_0(T)} \hat{b}^+_{\vec{k}} \hat{b}_{\vec{k}} \tag{39}$$

### B. The wavefunction of the superconducting state

To begin, let us consider first the wavefunction in the representation of number of occupation for the most general case, in which the pair occupies a discrete state, which corresponds to a different temperature. In this case the wavefunction of the pair can be written as

$$\varphi(n, k, T) = |n_0 k_0 T_0, n_1 k_1 T_1, n_2 k_2 T_2, \ldots\ldots, n_{max} k_{max} T_C \rangle \tag{40}$$

The wavefunction for the ground state is given by

$$\varphi(n_0, 0, 0) = |n_0 00, 000, 000, \ldots\ldots, 000 \rangle \tag{41}$$

and for the critical temperature we have

$$\varphi(n_{max}, k_{max}, T_C) = |000, 000, \ldots\ldots, n_{max}, k_{max}, T_C \rangle \tag{42}$$

where $n_0, k_0, T_0$ characterize the ground state and only one of the states $n_k, k_k, T_k$ can be occupied by the pair. As it is known, the pairs do not interact to each other, for what it the wavefunction coincides with a plane wave and can be written as

$$\varphi \approx e^{i\vec{k}.\vec{r}} \tag{43}$$

where the value of $k$ is

$$k = \frac{\sqrt{\mu}}{\hbar}\left(\varepsilon_k^2 + \varepsilon_0^2\right)^{1/4} \tag{44}$$

then

$$\varphi \approx e^{\frac{i}{\hbar}\sqrt{\mu}\left(\varepsilon_k^2 + \varepsilon_0^2\right)^{1/4}} \tag{45}$$

For the ground state $\varepsilon_k = 0$ and the temperature $T = T_C$, besides the bounding energy is zero ($\varepsilon_0 = 0$).

### C. Dispersion law for the pairs

All the fundamental magnitudes that characterize a superconductor are functions of the temperature and of the critical temperature; therefore, it is necessary to identify the true functionality of each term in the Eq. (14).

Thus, $E(k)$ represents the maximum energy and is dependent of maximum reduced wavevector $k_{max}$, or it only depends on the critical temperature. $\varepsilon(k)$ depends on electron wavevector $k$ and it in turn depends on the temperature, and $\varepsilon_0$ is only function of the temperature. $\varepsilon$ and $\varepsilon_0$ are physical magnitudes varying in opposite directions, this is, when $\varepsilon$ is maximum then $\varepsilon_0$ is minimum and vice versa. Both vary so that the sum of their squares always is constant.

The problem is now to find the explicit dependence of each magnitude. For this purpose we will take into account the following: any change of the total energy beginning from the critical temperature is due only to the change of thermal and binding energies. Let also suppose that the energy states that are occupied by the pairs between 0 and $T_C$ are discrete or continuous. Then, in the discrete case exists a maximum number $n$ for which the energy of the pair is very near to the thermal energy corresponding to the critical temperature (energy of breakup of the superconducting state).

Under assumption that the variation of the energy below the critical temperature is quantized, we then can write that

$$E = \frac{1}{2}\left(n_{max} + \frac{1}{2}\right)k_B T_C \tag{46}$$

where $n_{max}$ is the maximum number of energy quantum of available energy gap $2\varepsilon_0$.

In the case when the pairs take the energy in discrete form, we have

$$E = \lambda k_B T_C \tag{47}$$

where $\lambda$ is a constant that depends of characteristics of the superconductors material.

Let us consider now the $\varepsilon$ energy. This energy for the temperature $T$ in the discrete case can be written in form

$$\varepsilon = \frac{k_B T}{2}\left(n + \frac{1}{2}\right) \tag{48}$$

where $n = 1, 2, 3, \ldots, n_{max}$

In the continuous case, we can write this energy in the same way as Eq. (47)

$$\varepsilon = \lambda k_B T \tag{49}$$

The general equation of the superconductor pairs is given by

$$\left[\frac{1}{2}\left(n_{max} + \frac{1}{2}\right)k_B T_C\right] = \sqrt{\left[\frac{k_B T}{2}\left(n + \frac{1}{2}\right)\right]^2 + \varepsilon_0^2(T)} \tag{50}$$

for the discrete case, but the continuous case is given by

$$\lambda k_B T_C = \sqrt{(\lambda k_B T)^2 + \varepsilon_0^2(T)} \tag{51}$$

### D. Energy gap

Starting from these equations the characteristic magnitudes of the superconductors can be determined as follows: the variation of the gap energy with the temperature, the variation of the critical magnetic field with the temperature, the energy of the ground state and any of the intermediates states, [etc].

From the Fig. 2 and the Eqs. (50) and (51) we can conclude that there can not exist the pairs with the energy of Fermi when the temperature is equal to zero, because at zero Kelvin the oscillations of the pairs can not be realized. Therefore the pairs don't possess condensation of Bose-Einstein because the energy doesn't goes to zero when the temperature goes to zero.

The Eqs. (50) and (51) are the general equations and they describe the properties of any metallic superconductors, where the electrical resistance is given by the electron-phonon interaction; it is only necessary to determine the $n_{max}$ through experimental measurements of the critical temperature and of the bound energy of the pair $2\varepsilon_0$.

In order to calculate the variation of the energy gap with the temperature, we start from the Eqs. (50) and (51) and we obtain the following equation

$$\varepsilon_0(T) = \left[\frac{1}{2}\left(n_{max} + \frac{1}{2}\right)k_B T_C\right]\left\{1 - \left[\frac{\left(n + \frac{1}{2}\right)}{\left(n_{max} + \frac{1}{2}\right)}\frac{T}{T_C}\right]^2\right\}^{1/2} \tag{52}$$

for the discrete case and

$$\varepsilon_0(T) = \lambda k_B T_C \left\{1 - \left(\frac{T}{T_C}\right)^2\right\}^{1/2} \tag{53}$$

for the continuous case.

In case of $T = 0K$ we have

$$\varepsilon_0(0) = \frac{1}{2}\left(n_{max} + \frac{1}{2}\right)k_B T_C \qquad (54)$$

for the discrete case and

$$\varepsilon_0(0) = \lambda k_B T_C \qquad (55)$$

for the continuous case.

Notice that $\lambda$ and $1/2(n_{max} + 1/2)$ are constants. They can be obtained from experimental measurements or theoretically determined with aid of the particular characteristics of the superconductor, keeping in mind the relationship of the critical temperature given by condition $H_{ee} + H_{eP} = 0$

In the case of $T = T_C$, $n = n_{max}$ and $\varepsilon_0(T_C) = 0$

### E. Critical magnetic field

The critical magnetic field in $T = 0K$ is related with $\varepsilon_0(0)$. But the energy gap in $T = 0K$ is $2\varepsilon_0(0)$, then

$$\frac{H_0^2}{8\pi} = N(0)\frac{\varepsilon_0^2(0)}{2} \quad \text{and} \quad H_0 = [4\pi N(0)]^{1/2}\frac{1}{2}\left(n_{max} + \frac{1}{2}\right)k_B T_C \qquad (56)$$

for the discrete case and for the continuous case we obtain

$$H_0 = [4\pi N(0)]^{1/2}\lambda k_B T_C \qquad (57)$$

The variation of the critical magnetic field with the temperature is related directly with $2\varepsilon_0(T)$, therefore

$$H_c^2 = H_0^2\left\{1 - \left[\frac{\left(n+\frac{1}{2}\right)}{\left(n_{max}+\frac{1}{2}\right)}\frac{T}{T_C}\right]^2\right\} \qquad (58)$$

or

$$H_c^2 = H_0^2\left\{1 - \left(\frac{T}{T_C}\right)^2\right\} \qquad (59)$$

### F. Entropy and specific heat

From the Fig. 1 one can see that one can consider the pair as a combination of two electrons with two holes, therefore one can use the expression for the entropy in terms of the distribution function $f$.

$$S = -k_B \sum_k \{f_{-k-q}\ln f_{-k-q} + (1-f_{-k})\ln(1-f_{-k}) + f_{k+q}\ln f_{k+q} + (1-f_k)\ln(1-f_k)\} \qquad (60)$$

or

$$S = -k_B \sum_k \{[f_{k+q}\ln f_{k+q} + (1-f_{-k})\ln(1-f_{-k})] + [f_{-k-q}\ln f_{-k-q} + (1-f_k)\ln(1-f_k)]\} \qquad (61)$$

being
$$\varepsilon_{k+q} = \sqrt{E_{Max}^2 + \varepsilon_0^2(T)}; \qquad \varepsilon_k = -\sqrt{E_{Max}^2 + \varepsilon_0^2(T)} \tag{62}$$
$$\varepsilon_{-k-q} = -\sqrt{E_{Max}^2 + \varepsilon_0^2(T)}; \qquad \varepsilon_{-k} = \sqrt{E_{Max}^2 + \varepsilon_0^2(T)} \tag{63}$$
then
$$\varepsilon_{k+q} = \varepsilon_{-k}; \qquad \varepsilon_{-k-q} = \varepsilon_k \tag{64}$$
The energy of the electron pairs is
$$\varepsilon_{k+q} - \varepsilon_{-k-q} = 2\sqrt{E_{Max}^2 + \varepsilon_0^2(T)}; \qquad \varepsilon_k - \varepsilon_{-k} = -2\sqrt{E_{Max}^2 + \varepsilon_0^2(T)} \tag{65}$$
where the electrons and holes obey the Fermi statistic
$$f_k = f_{-k-q} = \frac{1}{e^{\beta\varepsilon}+1} \quad \text{and} \quad f_{-k} = f_{k+q} = \frac{1}{e^{-\beta\varepsilon}+1} \tag{66}$$
with $\beta = 1/k_B T$
We also know that
$$\frac{f_{-k}}{1-f_{-k}} = e^{-\beta\varepsilon} \quad \text{and} \quad \ln\frac{f_{-k}}{1-f_{-k}} = -\beta\varepsilon \tag{67}$$
of the same way
$$\frac{f_k}{1-f_k} = e^{\beta\varepsilon} \quad \text{and} \quad \ln\frac{f_k}{1-f_k} = \beta\varepsilon \tag{68}$$
then
$$S = -2k_B \sum_k \left\{ \left[ f_{-k} \ln\frac{f_{-k}}{1-f_{-k}} + \ln(1-f_{-k}) \right] + \left[ f_k \ln\frac{f_k}{1-f_k} + \ln(1-f_k) \right] \right\} \tag{69}$$
or
$$S = -2k_B \sum_k \left\{ \left[-\beta\varepsilon f_{-k} - \ln(1+e^{-\beta\varepsilon})\right] + \left[\beta\varepsilon f_k - \ln(1+e^{\beta\varepsilon})\right] \right\} \tag{70}$$
The factor 2 in the entropy formulate is due to that the electron in $k+q$ is equivalent to the hole in $-k$ and the electron in $-k-q$ is equivalent to the hole in $k$.
Then
$$S = 4k_B \sum_k \left\{ \beta\varepsilon f_{-k} + \ln(1+e^{-\beta\varepsilon}) \right\} \tag{71}$$
The specific heat, then is given by the expression
$$C_S = -\beta\frac{dS}{d\beta} = -4\beta k_B \sum_k \left\{ \beta\varepsilon \frac{df_k}{d\beta} \right\} \tag{72}$$
Also the following expression for the entropy can be obtained
$$C_S = \beta^2 k_B \sum_k \left\{ f_{-k}(1-f_{-k})\left[\varepsilon^2 + \frac{\beta}{2}\frac{d\varepsilon^2}{d\beta}\right] + f_k(1-f_k)\left[\varepsilon^2 + \frac{\beta}{2}\frac{d\varepsilon^2}{d\beta}\right] \right\} \tag{73}$$
but
$$f_k(1-f_k) = f_{-k}(1-f_{-k}) \tag{74}$$
Finally, we have that the entropy is given by

$$C_S = 2\beta^2 k_B \sum_k \left\{ f_k(1-f_k)\left[\varepsilon^2 + \frac{\beta}{2}\frac{d\varepsilon^2}{d\beta}\right]\right\} \quad (75)$$

### G. Coherence distance $\xi_0$

From the Fig. 1 we can see that, for $T \approx 0K$
$$\varepsilon(k+q) - \varepsilon(k) = E_{max} = 2\varepsilon_0(0) \quad (76)$$
but
$$\varepsilon(k+q) - \varepsilon(k) = \frac{\hbar^2(k+q)^2}{2m^*} - \varepsilon_F - \frac{\hbar^2 k^2}{2m^*} + \varepsilon_F = \frac{\hbar^2 q^2}{2m^*} + \frac{\hbar^2}{m^*}kq = 2\varepsilon_0 \quad (77)$$
From the Eq. (4) we have that
$$q = 2(k_F - k) \quad (78)$$
then
$$2\frac{\hbar^2}{m^*}(k_F - k)^2 + 2\frac{\hbar^2}{m^*}k(k_F - k) = 2\varepsilon_0 \quad (79)$$
and
$$\frac{\hbar^2}{m^*}k_F(k_F - k) = \frac{\hbar^2}{2m^*}qk_F = \varepsilon_0 \quad (80)$$
taking into account that
$$v_F = \frac{\hbar k_F}{2m^*} \quad (81)$$
we finally obtain
$$q = \frac{\varepsilon_0}{\hbar v_F} \quad (82)$$
which means that
$$\xi_0 = \frac{1}{\pi q} \quad (83)$$
or, that
$$\frac{1}{\xi_0} = \frac{\pi \varepsilon_0}{\hbar v_F} \quad (84)$$

## VII. RESULTS OF THE BCS THEORY

The BCS theory[1] of the superconductivity is a theory that possesses many limitations and their formulation is not totally microscopic, since part of its foundation is empiric and phenomenological. In this theory the strong criteria of the superconductivity is confuse. The BCS theory[1] considers that the electron-phonon interaction is greater than the Coulomb repulsion and of opposite sign in such a way that their sum is considered as a constant negative potential $-V$. This negative potential doesn't appear in any of the important magnitudes of the superconductivity. Also it is not made a correct interpretation of the Eq. (14).

In order to obtain the results of the BCS theory of our general expressions we should fix the maximum $n$. In spite of that in BCS theory[1] the $n_{max} = 3$, with this unique value is explained the properties of the superconductors characterized by the critical temperature less than $4K$. However for superconductors of greater critical temperatures, like $Pb$, $Nb$, $V$, etc., the $n_{max} = 3$ gives serious discrepancy of the theory and the experimental data, therefore, it is necessary to introduce a maximum $n$ equal to 4. This last number is not considered by the BCS theory[1].

The relationship that determines the energy gap up to zero Kelvin degree is given by

$$\varepsilon_0(0) = 1.75 k_B T_C \quad (85)$$

This equation is equivalent to

$$2\varepsilon_0 = 3.5 k_B T_C \quad (86)$$

which it coincides with the Eq. (3.30) of the BCS theory[1].

The BCS theory behaves as the continuous case, where $\lambda = 1/2 (n_{max} + 1/2)$, that is to say, $n = n_{max}$. In the BCS theory isn't deduced an equation of variation of the energy gap with the temperature and instead of that it is used the equation suggested by Buckingham[2] Eq. (3.31)[1]. This formulation is empiric one and is only worth near the critical temperature. We obtain the following expression of our results for the BCS superconductors.

$$\varepsilon_0(T) = 1.75 k_B T_C \left[1 - \left(\frac{T}{T_C}\right)^2\right]^{1/2} \quad (87)$$

We should keep in mind that with $T = 0K$ and from the previous expression, the Eq. (87) is obtained. Also if $T = T_C$, $n = n_{max} = 3$ and $\varepsilon_0(T_C) = 0$ is obtained too.

### A. Critical magnetic field

The critical magnetic field is the field that has sufficiently energy to break the superconducting state, that is to say, the critical magnetic field is related with $\varepsilon_0^2(T)$. Therefore we can write the following relation

$$N(0)\left\{(1.75 k_B T_C)^2 - \left[\frac{k_B T}{2}\left(n_{max} + \frac{1}{2}\right)\right]^2\right\} = N(0)\varepsilon_0^2(T) = \frac{H_C^2(T)}{8\pi}, \quad (88)$$

where $N(0)$ is the density of states at $T = 0K$

Keeping in mind that the energy gap is $2\varepsilon_0(0)$, then

$$H_C(0) = 1.75(4\pi N(0))^{1/2} k_B T_C \quad (89)$$

This equation is the same that the Eq. (3.39) of the BCS theory[1].

The BCS theory doesn't obtain a correct relationship for the variation of the critical magnetic field with the temperature. Thus, the Eqs. (3.43), (3.44) is an approach that doesn't reproduce the result for $T = T_C$. However in the Fig. 2 of the BCS theory comes a result that isn't in accordance with the obtained expressions.

From our results we can derive a general expression for the critical magnetic field as a function of the temperature:

$$H_C^2(T) = H_0^2(0)\left[1 - \left(\frac{T}{T_C}\right)^2\right] \tag{90}$$

### B. Specific heat

The heating capacity in the superconducting state is given by

$$C_S = -\beta \frac{dS}{d\beta} = -4\beta k_B \sum_k \left\{\beta \varepsilon \frac{df_k}{d\beta}\right\} \tag{91}$$

At the transition temperature the energy gap vanishes and the jump in specific heat associated with the second order transition is given by

$$C_S - C_N = -4\beta k_B \sum_k \beta \varepsilon \frac{df_k}{d\beta} - \frac{\pi^2}{3} k_B^2 N(0) T \tag{92}$$

For $T = T_C$

$$C_S - C_N = -4k_B \beta_C^2 \int N(0)\varepsilon \frac{df_k}{d\varepsilon} \frac{d\varepsilon}{d\beta} d\varepsilon - \frac{\pi^2}{3} k_B^2 N(0) T_C \tag{93}$$

considering that

$$\varepsilon \frac{d\varepsilon}{d\beta} + \varepsilon_0(T) \frac{d\varepsilon_0(T)}{d\beta} = 0 \tag{94}$$

and knowing that our theory

$$\frac{d\varepsilon_0^2(T)}{d\beta} = 2\left[\frac{1}{2}\left(n_{max} + \frac{1}{2}\right)\right]^2 \frac{1}{\beta_C^3} \tag{95}$$

finally we obtain the following relation for the jump of the specific heat

$$\frac{C_S - C_N}{\gamma T_C} = 8 \frac{3}{\pi^2} \left[\frac{1}{2}\left(n_{max} + \frac{1}{2}\right)\right]^2 (f_0 - f_{\varepsilon_{max}}) - 1 \tag{96}$$

The value for superconductors BCS with $T_C < 3K$ is

$$\frac{C_S - C_N}{\gamma T_C} = 1.6210 \tag{97}$$

This value is better that the presented in the BCS theory[1], which is 1.52
A general expression for the specific heat is derived from the previous relations

$$\frac{C_S}{\gamma T_C} = 8 \frac{3}{\pi^2} \left[\frac{1}{2}\left(n_{max} + \frac{1}{2}\right)\right]^2 \left(\frac{T}{T_C}\right)^3 (f_0 - f_{\varepsilon_{max}}) \tag{98}$$

In the figure 4 the data for tin and vanadium are presented and are compared with the relation obtained in the BCS theory[1] and the experimental results. Our results describe very well the experimental curves in the whole interval of temperature. The BCS theory doesn't describe the experimental results near the critical temperature.

### VIII. CONCLUSIONS

We have presented a new theory of the superconductivity with a new approach in the formation of the pairs. Our pairs are formed with electrons that belong to different bands that are symmetrical with respect to the Fermi level. This leads to the opposite signs for the masses of the paired electrons (see Fig. 1). The condition for the existence of the superconductivity in this case is that the electrical resistance be zero, therefore the electron-phonon interaction should not exist, because it is responsible for the resistance. This hypothesis means that the electron-electron interaction is responsible for the compensation of the electron-phonon interaction. From this condition we conclude that the electron-electron interaction and the electron-phonon one depend on the temperature and are equal. From this equality the dependence of the critical temperature can be obtained as the characteristic parameter of the material. This condition also tells us when a material should be superconductor or not as well as how to increase the critical temperature of a superconductor material

The relation $2\varepsilon_0 = 3.5 k_B T_C$ is imposed in the BCS theory[1] from the empiric point of view and it forbids the existence of superconductivity of high critical temperature. We think that this relation is important and it should be written in the following form: $2\varepsilon_0 = \lambda k_B T_C$, where $\lambda$ is a parameter that can be discrete or continuous, but depends on the characteristics of the material. This is due to the fact that the critical temperature was obtained under assumption that the electron-electron and electron-phonon interactions are equal (the parameter defines which characteristic should have a material in order to be a good or bad superconductor). This parameter can be experimentally measured. For the superconductors characterized by high critical temperatures this parameter is big.

The ceramic superconductors can be treated as metallic ones due to the planes or chains of metallic origin. For this reason the electric resistance is given by the electron-phonon interaction. If $\lambda$ is determined theoretically or experimentally, then it will address the main properties of these superconductors.

In order to complete this work, it would be interesting to calculate the Landau levels for the case of a superconductor in an external magnetic field. If some of these levels fall inside the region of the energy gap, then the individual electronic states will coexist with the superconductor pairs. In such case alone would obtain the so- called superconductor type II. Work is in progress on this subject and it will be published elsewhere.

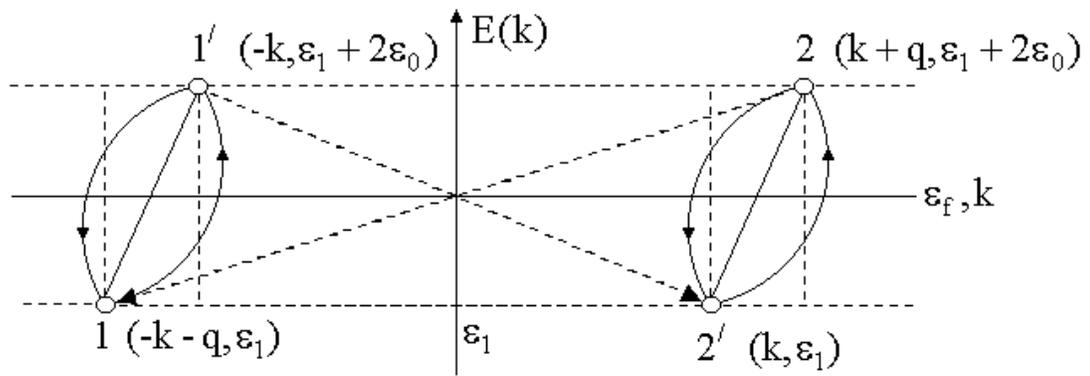

Fig. 1 Energy and states of the electron pairs coupled around the Fermi level.

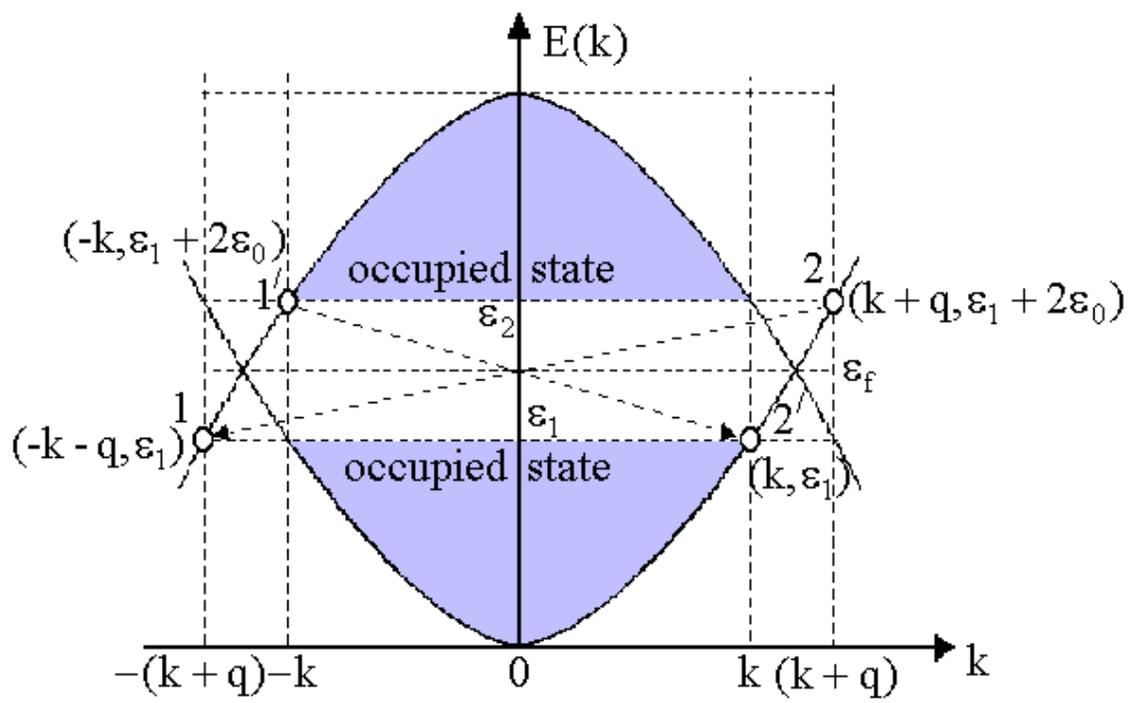

Fig. 2 Symmetry breaking of electron pair bands.

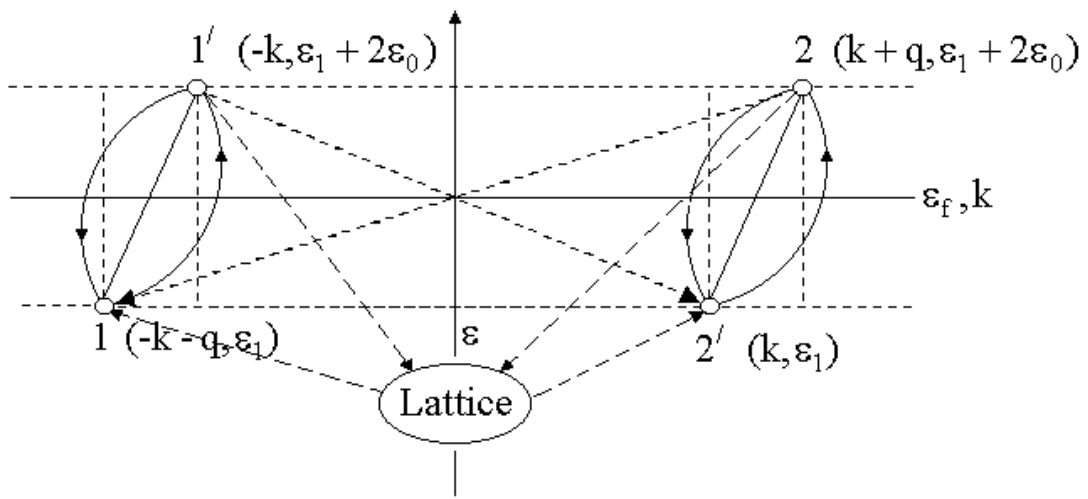

Fig. 3 Participation of the crystalline lattice in the process of formation of the electron pairs.

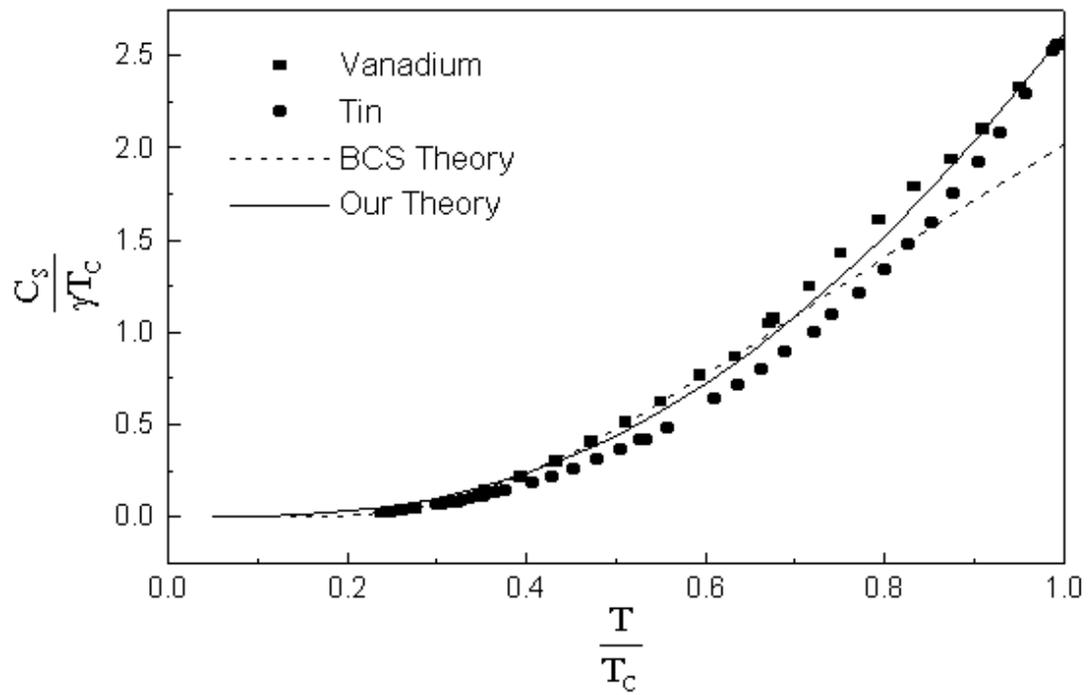

Fig. 4 Experimental and theoretical specific heats.